\documentstyle[12pt]{article}
\begin{document}
\begin{center}
{\large \bf 
Multifractal Measures in Iterative Maps\\

\vspace*{.5in}

\normalsize 
Kyungsik Kim$^{*}$, B. O. Shim and Y. S. Kong\\

\vspace*{.1in}

{\em Department of Physics, Pukyong National University,\\
Pusan 608-737, Korea}\\ 

\vspace*{.1in}

\hfill\\
B. I. Henry 
\vspace*{.1in}\\
{\em Department of Applied Mathematics, University of New South Wales,\\
Sydney NSW 2052, Australia}\\

\vspace*{.1in}

\hfill\\
M. K. Yum 
\vspace*{.1in}\\
{\em Department of Pediatric Cardiology, Hanyang University,\\
Kuri 471-701, Korea}
\vspace*{.1in}\\
\hfill\\
}
\vspace*{.3in}

%
\hfill\\
\end{center}
%
%
\baselineskip 24pt
\vskip 15mm     
We investigate chaotic and multi-fractal properties of a
two parameter map of the unit interval onto itself
-- the Kim-Kong map.
These results are compared with similar properties in well known
one parameter maps of the unit interval onto itself. 
\vskip 30mm
\noindent
$^{*}$E-mail: kskim@dolphin.pknu.ac.kr, Fax: +82-51-611-6357
\newpage
\noindent
\vskip 2mm
\indent
Over the past few decades, following the discovery of chaotic behaviour
in deterministic nonlinear dynamical systems, there has been
considerable interest in the study of
nonlinear differential equations and nonlinear difference
equations with applications ranging across
physics, chemistry, biology and economics.
In many of these studies a knowledge of the properties
of simple one-dimensional maps such
as the
logistic map, the dissipative standard map, the Lorentz map,
the  H\'enon map, the tent map, 
the quadratic map, etc.,
 has provided fundamental insights$[1-5]$. 
This has been particularly the case
in understanding deterministic chaos, understanding
various transitions to chaos as a control parameter is varied, and identifying
universality classes for the onset of chaos similar
to universality in the theory of
critical phenomena.

Most detailed studies of one-dimensional maps have been confined
to maps with a single control parameter
and many of the chaotic and universal
features of these maps are now well understood.
For example, Feigenbaum$[5]$ 
identified universal values for convergence ratios in the period-doubling route
to chaos in certain one-dimensional maps.
Universality has also been identified in the
quasiperiodic route to chaos and in intermittency$[3]$.
Two of the most useful quantitative measurements of dynamical properties
of single parameter one-dimensional maps have been
furnished by Lyapunov exponents and fractal dimensions.
In this paper we present the results of our measurements of these and
related quantities, multifractal spectrum,
for a one-dimensional map with two control parameters - the Kim-Kong map$[6]$. 
The motivation for this work is twofold;
firstly explore multifractal analysis
for discrete dynamical systems and secondly to investigate the
possibly competing effects of two control parameters
in discrete dynamical systems.

The majority of the applications of multifractal analysis$[7-9]$
to date have been
to physical systems.
For example in 
recent applications,
Kim and Kong used the box-counting method
to compute generalized dimensions and scaling exponents
for mountain heights$[10]$ 
and sea-bottom depths$[11]$ 
in Korea.

The Kim-Kong map is a two parameter map of the unit interval
 onto itself defined by$[6]$
\begin{eqnarray}
x(n+1)&=&f(x(n), \gamma, \beta )\nonumber\\
& = &\gamma\exp [-\beta (\log x(n))^2][1-\exp [-\beta (\log x(n))^2]]
\label{eq:a1}
\end{eqnarray}                             
with $x(n)\in (0,1)$, where $\gamma$ and $\beta$ are the
two control parameters.
To begin with we note that
the Kim-Kong map is a dissipative unimodal map with a locally 
quadratic maximum. The maximum value is $f(x)=\frac{\gamma}{4}$ which
occurs at $x= \exp(-\sqrt{\frac{\ln 2}{\beta}})$. A series expansion about 
the maximum yields
\begin{equation}
f(x) \approx \frac{\gamma}{4} - \gamma\beta \exp(-2 \sqrt{\frac{\ln 2}{\beta}})
(x-\exp(-2 \sqrt{\frac{\ln 2}{\beta}}))^2,
\label{eq:b2}
\end{equation} 
and the Schwarzian derivative is negative for all $x \in (0,1)$.     
If $\frac{\gamma}{4}>\exp(-\sqrt{\frac{\ln 2}{\beta}})$ then the map
has two fixed points $x_1$ and $x_2$ on the interval $(0,1)$ with one on either
side of the maximum, i.e., $0<x_1<\exp(-\sqrt{\frac{\ln 2}{\beta}})<x_2<1$.
The map also has a fixed point at $x=0$ and while this is outside
the domain $(0,1)$ it is an attracting fixed point for
$x\in(0,x_1)$. Numerical underflow can be avoided by selecting parameter values
$\gamma$ and $\beta$ together with the initial point $x(0)$ so that
the twp conditions $x_1<f(\frac{\gamma}{4})$ and $x_1<x(0)<\frac{\gamma}{4}$
are satisfied. Future iterates will
the be confined to $x\in(x_1,\frac{\gamma}{4})$.

The function $f(x)$ is displayed in
Fig.1 for
$\gamma=3.78$ and several values of $\beta=0.2,$ $0.6,$ $1.0,$ $1.5,$ $2.0$.
From this figure it can be seen that the main effect of
increasing $\beta$ at a fixed value of $\gamma$ is to
shift the position of the maximum to higher values of $x$.
In contrast to the logistic map, the Kim-Kong map
is not symmetric about the maximum.
In Fig. 2 we have displayed
an empirical unimodal one-dimensional map that was obtained from 
Belousov-Zhabotinskii reaction experiments$[12-14]$.
By comparing Figs. 1 and 2 it can be seen that
the functional form of the Kim-Kong map
for $0<\beta<0.2$ and $\gamma=3.78$ is similar
to the experimentally obtained map.

The Kim-Kong map undergoes a period doubling cascade
to chaos at fixed $\beta$ as $\gamma$ is increased.
An example of the period doubling cascade is shown in
in Fig. 2 of $[6]$
for $\beta=0.2$ and $0<\gamma<3.78$.
Denoting $\gamma_k$ as the parameter at which the period $2^{k-1}$
cycle becomes unstable we compute the sequence
$$
\delta_k=\frac{\gamma_{k+1}-\gamma_k}{\gamma_{k+2}-\gamma_{k+1}}
$$
from which we deduce that
$\delta = \lim_{k\rightarrow \infty} {\delta}_k\approx 4.66920$ 
in agreement with Feigenbaum's universal scaling result$[5]$
for dissipative uni-modal maps. The other scaling exponent,
$\alpha\approx 2.50290$ for successive branching
widths is also recovered in the Kim-Kong map.

We now consider a multifractal analysis$[7-9]$ 
of chaotic
orbits of the Kim-Kong map.
First we review the definition of generalized dimensions
in the multifractal formalism.
Suppose that we divide the unit interval into $M(\epsilon)$
cells of size $\epsilon=\frac{1}{M(\epsilon)}$.
For a given chaotic orbit of length $N$  let $p_i=\frac{n_i}{N}$ denote the
fractional number of points in the $i$th cell, $[(i-1)\epsilon,(i)\epsilon]$.
The generalized dimensions are now defined by
\begin{equation}
D_q = \lim_{\epsilon\rightarrow 0}
\frac{\log\sum_{i=1}^{M(\epsilon)}p_i^q}{(q-1)\log\epsilon}.
\label{eq:d14}
\end{equation}
and the multifractal measures
$f_q$ and $\alpha_q$ are related
to $D_q$ via
the Legendre transform: 
\begin{equation}
f_q = q \frac{d}{dq}[(q-1)D_q]-(q-1)D_q 
\label{eq:e15}
\end{equation}
and
\begin{equation}
\alpha_q = \frac{d}{dq}[(q-1)D_q]. 
\label{eq:f16}
\end{equation}
We have made use of Eqs.(3)-(5) to compute 
the multifractal measures for the Kim-Kong map
and the
fractal dimension, $D_0$, 
which is numerically compared with that of other maps.

In our numerical computations we
have restricted ourselves to
 three cases for the two control parameters in the chaotic regime;
(a) $\gamma=3.0$ and $\beta=0.12$, (b) $\gamma=3.4$ 
and $\beta=0.12$, and (c) $\gamma=3.5$ and $\beta=0.12$.
The multifractal measures that we have computed are based
on $2\times10^4$ iterations of the map in Eq.(1).
Figs. 4 and 5 are, respectively, the plots of $D_q$ versus $q$ and the spectrum 
$f_q$ versus $\alpha_q$ for chaotic orbits in the Kim-Kong map. 
The fractal dimension $D_0$ or $f_0$ ( i.e., the maximum value of 
$D_q$ or $f_q$ ) of the Kim-Kong map is numerically
 compared with that of other maps, 
in Table 1.

In conclusion, we have computed chaotic and multifractal
properties of the Kim-Kong map - a  two-parameter map
of the unit interval onto itself.
Although this map was investigated as a mathematical
example in this work the map could prove useful
as a model for
the transport of quantum excitations, directed polymer problems,
and the electron localization in quantum mechanics$[18-20]$. 
In future work on the Kim-Kong map we plan to
investigate  chaotic orbits of this map as
examples of deterministic diffusion including the possibility of biased
 diffusion and anomalous diffusion. 
The chaotic orbits of the Kim-Kong maps will also
be compared with
randomness 
in a reaction-diffusion system with multi-species reactants$[21-22]$.
 \\  
\vskip 10mm
\noindent
{\bf ACKNOWLEDGMENT}
\vskip 3mm
\hfill\\
This work was supported by grant No.2000-2-133300-001-3 from the Basic Research 
Program of the Korea Science and Engineering Foundation.
\vspace {10mm}
%

%
\newpage
\vskip 10mm
%
{\bf FIGURE  CAPTIONS}
\vspace {10mm}

\noindent
Fig. 1. A plot of $x(n+1)$ verse $x(n)$ graph in the Kim-Kong map$[6]$,
$x(n+1) = 
\gamma \exp [-\beta (\log x(n))^2][1-\exp [-\beta (\log x(n))^2]]$,
for $\gamma=3.78$ and several values of
$\beta=0.2,$ $0.6,$ $1.0,$ $1.5,$ $2.0$. 
\vspace {10mm}

\noindent
Fig. 2.  Plots of two types of maps for chaotic states with impure malonic
acid in the Belousov-Zhabotinskii reaction[12], where
the circle and triangle data are the different U sequence of periodic states,
respectively.
   
%

%
\vspace {10mm}

\noindent
Fig. $3$.  Plots of $D_q$ and $q$
for $\gamma=3.0$ and $\beta=0.12$ (thick solid line), $\gamma=3.4$ 
and $\beta=0.12$ (dashed line),
 and $\gamma=3.5$ and $\beta=0.12$ (thick dashed line).
\vspace {10mm}

\noindent
Fig. $4$.  Plots of $f_q$ versus $\alpha_q$ obtained
 from the chaotic sequences of
the Kim-Kong map for three cases of the parameters;
 (a) $\gamma=3.0$ and $\beta=0.12$ (thick solid line), 
(b) $\gamma=3.4$ and $\beta=0.12$ (dashed line), and
 (c) $\gamma=3.5$ and $\beta=0.12$ (thick dashed line).
\vspace {20mm}

%
{\bf TABLE  CAPTIONS}
\vspace {10mm}

\noindent
Table $1$.  Summary of values of the fractal dimensions
of chaotic orbits
in the Kim-Kong map and other maps.

\newpage
\vspace {10mm}
\begin{tabular}{lcr} \hline\hline
Maps & Control Parameter & Fractal Dimension  \\  \hline
Kim-Kong Map            & (a) $\gamma=3.0, \beta=0.12$ & $0.8395$\\
                & (b) $\gamma=3.4, \beta=0.12$ & $0.9790$\\
              & (c) $\gamma=3.5, \beta=0.12$ & $0.9951$\\ 
H$\acute{e}$non map$[15]$ & $a=1.4,b=0.3$ & $1.26$\\
Kaplan-Yorke Map$[15]$ & $\alpha=0.2$ & $1.431$\\
Logistic Map$[16]$ & $b=3.5699456...$ & $0.538$\\
Lorentz Equation$[15,17]$ &   & $2.06\pm0.01$\\
Zaslavskii Map$[16]$ &     & $1.39$\\  \hline\hline
\end{tabular}

\end{document}